\documentclass[runningheads]{llncs}
\usepackage[T1]{fontenc}
\usepackage{graphicx}
\usepackage{hyperref}
\begin{document}
\title{Research Software Publication Policy Case Study}
%
%
\author{Nic Weber \inst{}}
\authorrunning{Weber}
%
\institute{University of Washington
\email{nmweber@uw.edu}}
\maketitle              
\begin{abstract}
Research software is increasingly recognized as a vital component of the scholarly record. Journals offer authors the opportunity to publish research software papers, but often have different requirements for how these publications should be structured and how code should be verified. In this short case study we gather data from 20 Physical Science journals to trace the frequency, quality control, and publishing criteria for software papers. Our goal with the case study is provide a proof-of-concept for doing descriptive empirical work with software publication policies across numerous domains of science and engineering. In the narrative we therefore provide descriptive statistics showing how these journals differ in criteria required for archiving, linking, verifying, and documenting software as part of a formal publication. The contribution of this preliminary work is twofold: 1. We provide case study of Physical Science research software publications over time; 2. We demonstrate the use of a new survey method for analyzing research software publication policies. In our conclusion, we describe how comparative research into software publication policies can provide better criteria and requirements for an emerging software publication landscape. 

\keywords{Research Software, Scholarly Communications}
\end{abstract}
\section{Introduction}

Research software is often developed by researchers for the specific purpose of transforming, analyzing, and visualizing data, and is often necessary for producing novel results in a computational domain. Despite the important role of software development in scientific and engineering research these contributions are rarely recognized as unique and credit-worthy contributions to knowledge production \cite{carver_conceptualization_2018,weber_finite_2020-1}. Multiple surveys have reported that researchers feel their work would be impossible to conduct without time, access, and money to develop research software \cite{hettrick2014uk}. However, research publications that present novel results remain the key indicator of credit and acknowledgment of scholarly contributions. Previous work has described why this model of research publication is incompatible with a current ecosystem where significant effort is devoted to developing software tools in order to produce novel research results \cite{anzt_environment_2021,niemeyer_challenge_2016}. Previous scientometric work has also shown that software is often cited in traditional research publications \cite{li_software_2016,park_research_2019}, but tracing and identifying these software mentions is difficult due to disparate practices in publishing and archiving research software \cite{howison_software_2016,weber}. 

Software papers can help improve the recognition and reward of development activities that are critical to advancing science and engineering. In a research software paper, the main focus is describing and verifying software used for a specific research task. This includes providing a description of the software’s intended use, its architecture and design, providing documentation, and a proof of concept for how the software can be used \cite{katz_recognizing_2021}). Scholarly journals have recently started to give authors the ability to publish research software papers, but practices across research domains vary substantially in how these papers are to be structured, and what information they must contain in order to be accepted for publication. For example, the journal ‘Astronomy and Computing’ will accept software-focused papers if they describe “outcomes (positive and negative) of the practical application of informatics techniques within astronomy” \cite{pasian_astronomy_2022}. Similarly, in Political Science the ‘Journal of Information Technology \& Politics’ offers a “workbench” paper where authors can describe research software in detail, but only if the author’s also include preliminary results using the software \cite{noauthor_journal_2022}. Dedicated journals, such as the Journal of Open Source Software (JOSS), SoftwareX, and the Journal of Open Research Software (JORS) have also attempted to create a lightweight model for software papers where authors can provide a general description of code, testing, and verification of a software contribution. 

Despite a growing practice of research paper publishing and analysis of software citation, there has been significantly less attention paid to how domain practices in software paper publications differ between journals. In this short paper we seek to, generally, characterize the software publication landscape through a survey of journal policies. We focus on software publication policies and practices in the broad domain of Physical Sciences in order to provide a proof of concept for how a large-scale survey of software paper policies can be conducted in the future. In the following sections we describe our methods of data collection and analysis, and then present results from analyzing 20 different journal policies related to software publications. 

\section{Methods}
To construct a sample of journal policies to review we first consulted a list of journals that publish software papers from the Software Sustainability Institute in the UK \cite{hong_which_2014}. From this initial list we then used the Web of Science’s journal subject-clustering tool to identify to create a domain identity for each journal. Here, we focus on the domain of Physical Sciences which includes subject-clusters such as “Earth and Planetary Sciences”, “Physics”, and “Astronomy”, and “Chemistry.” This resulted in 24 Physical Science journals. We then manually reviewed each journal’s policy about software publication and eliminated 4 journals that do explicitly mention they do not accept software-focused papers. Our final list of 20 journals can be found in the supplemental data to this paper. We then modified an evaluation framework from Candela et al that focused on surveying Data Journal policies \cite{candela_data_2015} to be applicable to research software.  In the following section, we explain each component of this modified framework and how it relates to a software publication policy. 

\subsection{Analysis} 
We first read all 20 journal’s publication policies, including the scope, directions to authors, and criteria for acceptance. We then manually recorded the nature of the publication, the instructions for a software paper’s preparation, whether the journal offered open access publication, and 10 criteria for how the software were to be archived and made available in the future. Below, we explain each area of analysis, and our results from analyzing 20 Physical Science journals. 

\subsubsection{Nature}
The nature of a journal is whether it only accepts software papers or a broader mix of research publications. JOSS, as described above, would be a “Pure” software journal because it only accepts research software publications, whereas ‘Astronomy and Computing’ would be a “Mixed” journal because it publishes research-driven papers as well as software papers. \textbf{Results}: All but one of the journals in the Physical Science sample were Mixed journals. “Computing and Software for Big Science” (Springer) was the only pure software journal that we identified in this domain that was devoted purely to software publications. 

\subsubsection{Length}
We characterized each journal’s requirements for page length for a software publication in order to better understand acceptance criteria for a software paper’s contribution. We determined the length parameter by searching for explicitly stated limits in Author Instructions. If no explicit guidance was given we then looked at the lengths of five articles focused explicitly on software. If all articles were four or fewer pages, the journal was listed as having a “Short” length requirement, but if any were more than four pages, the journal was listed as “Normal”. \textbf{Results}: The journal “Computational Molecular Science” (Wiley) was the only journal that contained “short” software papers. All of the remaining journals in our sample neither set nor enforced a length restriction on software papers. 

\subsubsection{Access status}
We recorded whether a journal was default open-access, or was not open-access but allowed for authors to publish their work with an open-access license for a fee. \textbf{Results}: 17 of the journals in our Physical Sciences domain cluster offered open-access publishing for a fee. Only three journals were by-default licensed as open-access, these include “Journal of Cheminformatics” (Springer); “Geoscientific Model Development” (EGU); and, “Computational Astrophysics and Cosmology” (Springer). 

\subsubsection{Software paper requirements}
Next, we investigate requirements for publishing a software paper in a Physical Science journal. We first read through the “Author Instructions” of each journal and came up with an overarching classification where a journal either provided explicit, implicit, or no directions on how to prepare a software paper for publication. An implicit set of requirements stated that there was a unique structure that a software paper should follow, but otherwise had no requirements for how the software was to be evaluated or described in the paper.\\ 

\textbf{Results}: Four journals in our sample had an implicit set of requirements for a software paper, and four had no requirements or structure for a software paper. The remaining 12 journals all described explicit requirements for a software paper to be accepted for publication. For the journals with explicit requirements, we then created a set of categories and judged the strength of these requirements on a three point scale, where 3 represents a category that was required for publication, a 2 represents a suggested category, and 1 represents the categorical requirement was not present in the journal’s policy for publishing a research software paper. The categories that were documented for each journal include the following: 
\begin{itemize}
\item \textbf{Archiving}: Software papers are required to archive the software in an openly-accessible repository, such as on Github, Zenodo, or Dataverse; 
\item \textbf{Persistent link:} The authors must obtain a persistent link (e.g. a DOI) to the software repository where the software was stored; 
\item \textbf{License: }The software being described in the paper must have an explicit license governing its reuse or distribution; 
\item \textbf{Functionality}: The paper required software authors to describe the intended functionality of the software.
\item \textbf{Quality and Testing}: The paper should describe how software tests (e.g. Unit, Integration, Performance testing) or quality control was carried out on the software. 
\item \textbf{Dependencies}: If the software required the installation or use of other libraries the software dependencies must be communicated to readers.
\item \textbf{Documentation}: The software being described must include documentation for configuring, running, and using the software. 
\item \textbf{Application}: The software must provide a use case, or demonstration of use through an application to a research problem. 
\item \textbf{Future Development}: Authors are required to describe goals of future development, or how users can contribute to ongoing development efforts. 
\end{itemize}

In the following plot, we show the frequency distribution of each category (ranked on a scale of 3-1 as described above). It is worth noting that no journal in our sample had a requirement for each category that we analyzed. The most frequently required aspects of a software paper were that dependencies were documented, documentation was provided for future use, and a license governing distribution was described in the paper. Somewhat surprisingly more than half of our overall sample require software to be archived in order to publish a software paper. Perhaps this is an assumption that does not need to be formally documented or required. However, we caution that without explicit directions to authors there is the possibility that journals are publishing papers that describe software, but don’t actually provide access to that software in meaningful ways that would enable their reuse in research. This is also reinforced by the fact that more than half of our sample had no requirements for documenting future development plans, or how the scientific community could make contributions to a research software package.\\

\begin{figure}[h]
\includegraphics[scale=.75]{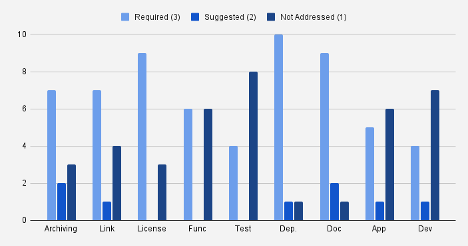}
\caption{Frequency distribution for software paper requirements across 20 journals in Physical Science}
\end{figure}

  \subsubsection{Software Paper Publishing Trends}
Next, we sought to understand the frequency of software papers published across all domains of research. In order to obtain this data, we again rely upon Hong’s list of journals that regularly publish research software papers \cite{hong_which_2014}, but in most of the journals that we analyzed there was no clear distinction between a software paper and a research paper. For journals that did not explicitly label a software publication, we read the abstract of each paper, and then judged whether the author’s indicated the primary contribution of the paper was about software, or were the results of using a specific piece of software. Below, we plot the publication frequency of software papers published annually from 2014-2021.

\begin{figure}[h]
\includegraphics[scale=.7]{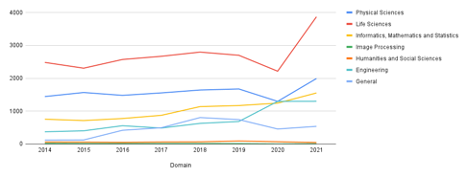}
\caption{Frequency distribution for software papers published from 2014-2021}
\end{figure}

As Figure 2 shows, there is a slight an upward trend in the number of research software papers published during this seven-year window. Informatics, Engineering, and Physical and Life Science journals have increasingly published software papers (absent a decline in overall publications during 2020). 

\section{Conclusions \& Future Work }
In this short paper we have described trends in software papers published in 20 Physical Science journals. We demonstrate that journals vary widely in their requirements for publishing a software paper, with most journals suggesting that dependency information and documentation are necessary components of a software publication. Further, we demonstrate an increasing tendency for many domains to publish research software papers. As we note in the findings described above it is curious that a minimum standard for archiving software is absent in over half of the journal policies that we analyzed. In future work, we will extend this survey method to analyze requirements across additional domains of science, engineering, and the humanities. This will help to substantiate the preliminary findings of this paper and allow for cross-domain comparisons of software publication trends.  The goal of our work, overall, is to rigorously describe an emerging software publication landscape, to recommend best practices in creating author instructions for research software papers, and to support ongoing methods development in measuring the impact of scientific software development. 

\subsubsection*{DATA ACCESSIBILITY}
Data behind this paper are available with a CC0 license in \href{https://doi.org/10.7910/DVN/DGKBLF}{Dataverse}. For more information on the broader project and software used to conduct this analysis please see our project's \href{https://github.com/BITS-Research/SoftwareJournalPolicySurvey  }{Github Repository}.

%
%
%
\bibliographystyle{splncs04}
\bibliography{ref}
\end{document}